# An Architectural Style for Ajax

Ali Mesbah and Arie van Deursen

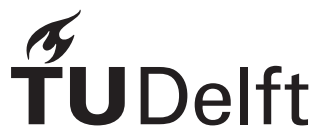
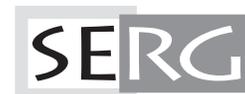

# An Architectural Style for Ajax

Ali Mesbah

*Software Evolution Research Laboratory*

*Delft University of Technology*

*Mekelweg 4, 2628 CD Delft, The Netherlands*

`A.Mesbah@tudelft.nl`

Arie van Deursen

*Software Engineering Group*

*Delft Univ. of Technology and CWI*

*Mekelweg 4, 2628 CD Delft, The Netherlands*

`Arie.vanDeursen@tudelft.nl`

**Abstract**

*A new breed of web application, dubbed* AJAX*, is emerging in response to a limited degree of interactivity in large-grain stateless Web interactions. At the heart of this new approach lies a single page interaction model that facilitates rich interactivity. We have studied and experimented with several* AJAX *frameworks trying to understand their architectural properties. In this paper, we summarize three of these frameworks and examine their properties and introduce the* SPIAR *architectural style. We describe the guiding software engineering principles and the constraints chosen to induce the desired properties. The style emphasizes user interface component development, and intermediary delta-communication between client/server components, to improve user interactivity and ease of development. In addition, we use the concepts and principles to discuss various open issues in* AJAX *frameworks and application development.*

## 1. Introduction

Over the course of the past decade, the move from desktop applications towards web applications has gained much attention and acceptance. Within this movement, however, a great deal of user interactiveness has been lost. Classical web applications are based on a *multi page interface* model, in which interactions are based on a page-sequence paradigm. While simple and elegant in design for exchanging documents, this model has many limitations for developing modern web applications with user friendly human-computer interaction.

Recently, there has been a shift in the direction of web development. A new breed of web application, dubbed AJAX (Asynchronous JavaScript And XML) [14], is emerging in response to the limited degree of interactivity in large-grain stateless Web interactions. At the heart of this new approach lies a *single page interface* model that facilitates rich interactivity. In this model, changes are made to individual user interface components contained in a web page, as opposed to refreshing the entire page.

Thanks to the momentum of AJAX, single page interfaces have attracted a strong interest in the web application development community. After the name AJAX was coined in February 2005 [14], numerous frameworks[1] and libraries have appeared, many web applications have adopted one or more of the ideas underpinning AJAX, and an overwhelming number of articles in developer sites and professional magazines have appeared. Adopting AJAX-based techniques is a serious option not only for newly developed applications, but also for existing web sites if their user friendliness is inadequate.

A software engineer considering adopting AJAX, however, is faced with a number of challenges. What are the fundamental architectural differences between designing a legacy web application and an AJAX web application? What are the different characteristics of AJAX frameworks? What do these frameworks hide? Is there enough support for designing such applications? What problems can one expect during the development phase? Will there be some sort of convergence between the many different technologies? Which architectural elements will remain, and which ones will be replaced by more elegant solutions?

Addressing these questions calls for a more abstract perspective on AJAX web applications. Despite all the attention the technology is receiving in the web community, there is a lack of a coherent and precisely described set of architectural formalisms for AJAX enabled web applications. In this paper we explore whether concepts and principles as developed in the software architecture research community can be of help to answer such questions. In particular, we propose SPIAR, an architectural style for AJAX applications, and study to what extent this style can help in addressing our questions.

This paper is organized as follows. We start out, in Section 2 by exploring AJAX, studying three frameworks (Google's GWT, Backbase, and the open source Echo2) that have made substantially different design choices. Then, in Section 3, we survey existing architectural styles (such as the Representational State Transfer architectural style REST on which the World Wide Web is based [12]), and analyze their suitability for characterizing AJAX. Next, in Section 4, we propose SPIAR, describing the architectural properties, elements, and constraints of this style. Given SPIAR, in Section 5 we use its concepts and principles to discuss various open issues in AJAX frameworks and application development. We conclude with a summary of related work, contributions, and an outlook to future work.

---

[1] At the time of writing more than 150 frameworks are listed at `http://ajaxpatterns.org/Frameworks`.

## 2. Ajax Frameworks

### 2.1. Ajax

AJAX [14] is the name given to a set of modern web application development technologies, previously known as *Dynamic HTML (DHTML)* and *remote scripting*, to provide a more interactive web-based user interface.

As defined by Garrett [14], AJAX incorporates: standards-based presentation using XHTML and CSS, dynamic display and interaction using the Document Object Model, data interchange and manipulation, asynchronous data retrieval using XMLHttpRequest, and JavaScript binding everything together. This definition, however, only focuses on the client side of the web application setting.

AJAX is an approach to web application development utilizing a combination of established web technologies. It is the combination of these technologies that makes AJAX unique and powerful on the Web.

Even before the term AJAX was coined, its power was becoming evident by web applications such as Google Suggest and Google Map. Other well known examples are Flickr, Gmail, and the new version of Yahoo Mail. For more technical details of AJAX see [1, 9].

### 2.2. Frameworks

Web application developers have struggled constantly with the limits of the HTML page-sequence experience, and the complexities of client-side JavaScript programming to add some degree of dynamism to the user interface. Issues regarding cross-browser compatibility are, for instance, known to everyone who has built a real-world web application. The rich user interface (UI) experience AJAX promises comes at the price of facing all such problems. Developers are required to have advanced skills in a variety of Web technologies, if they are to build robust AJAX applications. Also, much effort has to be spent on testing these applications before going in production. This is where frameworks come to the rescue. At least many of them claim to.

Because of the momentum AJAX has gained, a vast number of frameworks are being developed. The importance of bringing order to this competitive chaotic world becomes evident when we learn that 'almost one new framework per day' is being added to the list of known frameworks[2].

We have studied and experimented with several AJAX frameworks trying to understand their architectural properties. We summarize three of these frameworks in this section. Our selection includes a widely used open source framework called Echo2, the web framework offered by Google called GWT, and the commercial package delivered by Backbase. All three frameworks are major players in the AJAX market, and their underlying technologies differ substantially.

**Echo2**

Echo2[3] is an open-source AJAX framework which allows the developer to create web applications using an object-oriented, UI component-based, and event-driven paradigm for Web development. Its Java *Application Framework* provides the APIs (UI components, property objects, and event/listeners) to represent and manage the state of an application and its user interface.

All functionality for rendering a component or communicating with the client browser is specifically assembled in a separate module called the *Web Rendering Engine*. The engine consists of a server-side portion (written in Java/J2EE) and a client-side portion (JavaScript). The client/server interaction protocol is hidden behind this module and as such, it is entirely decoupled from other modules. Echo2 has an *Update Manager* which is responsible for tracking updates to the user interface component model, and for processing input received from the rendering agent and communicating it to the components.

The *Echo2 Client Engine* runs in the client browser and provides a remote user interface to the server-side application. Its main activity is to synchronize client/server state when user operations occur on the interface.

A *ClientMessage* in XML format is used to transfer the client state changes to the server by explicitly stating the nature of the change and the corresponding component ID the change has taken place on. The server processes the ClientMessage, updating the component model to reflect the user's actions. Events are fired on interested listeners, possibly resulting in further changes to the server-side state of the application. The server responds by rendering a *ServerMessage* which is again an XML message containing directives to perform partial updates to the DOM representation on the client.

**GWT**

Google has a novel approach to implementing its AJAX framework, Google Web Framework (GWT)[4]. Just like Echo2, GWT facilitates the development of UIs in a fashion similar to AWT or Swing and comes with a library of widgets that can be used. The unique character of GWT lies in the way it renders the client-side UI. Instead of keeping the UI components on the server and communicating the state changes, GWT compiles all the Java UI components to JavaScript code (compile-time). Within the components the developer is allowed to use a subset of Java 1.4 API to implement needed functionality.

---

[2] http://ajaxpatterns.org/wiki/index.php?title=AJAXFrameworks

[3] Echo2 2.0.0, www.nextapp.com/platform/echo2/echo/. Their on line demo is worth looking at!

[4] http://code.google.com/webtoolkit/

GWT uses a small generic client engine and, using the compiler, all the UI functionality becomes available to the user on the client. This approach decreases round-trips to the server drastically. The server is only consulted if raw data is needed to populate the client-side UI components. This is carried out by making server calls to defined services. The services (which are not the same as Web Services) are implemented in Java and data is passed both ways over the network using serialization techniques.

**Backbase**

Backbase[5] is an Amsterdam-based company that provided one of the first commercial AJAX frameworks. The framework is still in continuous development, and in use by numerous customers world wide.

A key element of the Backbase framework is the Backbase Presentation Client. This a standards-based engine written in Javascript that runs in the web browser. It can be programmed via a declarative user interface language called BXML. BXML offers library of UI controls, a mechanism for attaching actions to them, as well as facilities for connecting to the server asynchronously.

The server side of the Backbase framework is formed by BJS, the Backbase Java Server. It is built on top of JavaServer Faces (JSF)[6], the new J2EE presentation architecture. JSF provides a user interface component-based framework following the model-view-controller pattern. The interaction in JSF is, however, based on the classical page sequence model, making integration in a single page framework non trivial.

Backbase Java Server provides its own set of UI components and extends the JSF framework to provide a single page interface implementation. Any Java class that offers getters and setters for its properties can be directly assigned to a UI component property. Developers can use the components declaratively (web-scripting) to build an AJAX application.

The framework renders each declared server-side UI component to a corresponding client-side (BXML) UI component, and keeps track of changes on both component trees for synchronization.

The state changes on the client are sent to the server on certain defined events. These can be action events like clicking a button, or value change events such as checking a radio button. The server translates these state changes and identifies the corresponding component(s) in the server component tree. After the required action, the server renders the changes to be responded to the engine again in BXML format.

---

[5] www.backbase.com
[6] JavaServer Faces Specification v1.1, http://java.sun.com/j2ee/javaserverfaces/

### 2.3. Features

While different in many ways, these frameworks share some common architectural characteristics. Generally, the goals of these frameworks can be summarized as follows:

- Hide the complexity of developing AJAX applications - which is a tedious, difficult, and error-prone task,
- Hide the incompatibilities between different web browsers and platforms,
- Hide the client/server communication complexities,
- All this to achieve rich interactivity and portability for end users, and ease of development for developers.

The frameworks achieve these goals by providing a library of user interface components and a development environment to create reusable custom components. The architectures have a well defined protocol for small interactions among known client/server components. Data needed to be transferred over the network is significantly reduced. This can result in faster response data transfers. Their architecture takes advantage of client side processing resulting in improved user interactivity, smaller number of round-trips, and a reduced web server load.

## 3. Architectural Styles

### 3.1. Terminology

In this paper we use the software architectural concepts and terminology as used by Fielding [11] which in turn is based on the work of Perry and Wolf [23]. Thus, a software architecture is defined [23] as a configuration of architectural elements — processing, connectors, and data — constrained in their relationships in order to achieve a desired set of architectural properties.

An architectural style, in turn, [11] is a coordinated set of architectural constraints that restricts the roles of architectural elements and the allowed relationships among those elements within any architecture that conforms to that style. An architectural style constrains both the design elements and the relationships among them [23] in such a way as to result in software systems with certain desired properties.

An architectural system can be composed of multiple styles and a style can be hybrids of other styles. Styles can be seen as reusable [20] common architectural patterns within different system architectures and hence the term *architectural pattern* is also used to describe the same concept [3].

### 3.2. Existing Styles

User interface applications generally make use of popular styles such as Module/View/Controler [17] to describe large scale architecture and, in more specific cases, styles like C2 [27] to rely on asynchronous notification of state changes and request messages between independent components.

Many different network-based architectural styles [11], such as client/server [24], n-tier [29], and Code on Demand, exist but in our view the most complete and appropriate style for the Web, thus far, is the REpresentational State Transfer (REST) [12].

REST emphasizes the abstraction of data and services as resources that can be requested by clients using the resource's name and address, specified as a Uniform Resource Locator (URL) [4]. The style inherits characteristics from a number of other styles such as client/server, pipe-and-filter, and distributed objects.

The style is a description of the main features of the Web architecture through architectural constraints which have contributed significantly to the success of the Web.

It revolves around five fundamental notions: a *resource* which can be anything that has identity, e.g., a document or image, the *representation of a resource* which is in the form of a media type, *synchronous request-response interaction* over HTTP to obtain or modify representations, a *web page* as an instance of the application state, and *engines* (e.g., browser, crawler) to move from one state to the next.

REST specifies a client-stateless-server architecture in which a series of proxies, caches, and filters can be used and each request is independent of the previous ones, inducing the property of scalability. It also emphasizes on a uniform interface between components constraining information to be transferred in a standardized form.

### 3.3. A Style for Ajax

AJAX applications can be seen as a hybrid of desktop and web applications, inheriting characteristics from both worlds. Can we reuse styles from these worlds?

User interface styles such as C2 are meant specifically for peer-to-peer environments and thus are not suitable for Web applications.

AJAX frameworks provide back-end services through UI components to the client in an event-driven style whereas REST provides resources. AJAX architectures are also not so easily captured in REST, due to the following differences:

- While REST is suited for large-grain hypermedia data transfers, because of its uniform interface constraint it is not optimal for small data interactions required in AJAX applications.

- REST focuses on a hyper-linked resource-based interaction in which the client requests a specific *resource*. In contrast, in AJAX applications the user interacts with the system much like in a desktop application, requesting a response to a specific *action*.

- All interactions for obtaining a resource's representation are performed through a synchronous request-response pair in REST. AJAX applications, however, require a model for asynchronous communication.

- REST explicitly constrains the server to be stateless, i.e., each request from the client must contain all the information necessary for the server to understand the request. While this constraint can improve scalability, the tradeoffs with respect to network performance and user interactivity are of greater importance when designing an AJAX architecture.

Because of these requirement mismatches, we do not see how existing styles such as REST or C2 can help to address some of the questions raised in the introduction. Therefore, we will propose a style specifically tailored towards AJAX applications, and study if this style can be used for this purpose instead.

## 4. SPIAR Architectural Style

In this section, we first focus on the essential architectural properties of AJAX frameworks. Then the common architectural elements are presented and finally the constraints on those elements to achieve the properties are discussed. Our SPIAR architectural style describes the essence of what AJAX frameworks hide and prescribes constraints that such frameworks should adhere to. The style can be used when high user interaction and responsiveness is desired in web applications.

### 4.1. Architectural Properties

Below we discuss a number of architectural properties that relate to the essence of AJAX. Other properties, such as extensibility or security, that may be desirable for any system but are less directly affected by a decision to adopt AJAX, are not taken into account. Note that some of the properties discussed below are related to each other: for instance, user interactivity is influenced by user-perceived latency, which in turn is affected by network performance.

**User Interactivity**
Human-computer interaction literature defines interactivity as the degree to which participants in a communication process have control over, and can exchange roles in their mutual discourse. User interactivity is closely related to *usability* [13], the term used in software architecture literature. Teo *et al.* [28] provide a thorough study of user interactivity on commercial web applications. Their results suggest that an increased level of interactivity has positive effects on user's perceived satisfaction, effectiveness, efficiency, value, and overall attitude towards a Web site. Improving this property on the Web has been the main motivating force behind the AJAX movement.

**User-perceived Latency**

User-perceived latency is defined as the period between the moment a user issues a request and the first indication of a response from the system. Generally, there are two primary ways to improve user-perceived performance. First, by reducing the round-trip time and second, by allowing the user to interact asynchronously with the system. This is an important property in all distributed applications with a front-end to the user.

**Network Performance**

Network performance is influenced by *throughput* which is the rate of data transmitted on the network and *bandwidth*, i.e., a measure of the maximum available throughput. Network performance can be improved by means of reducing the amount and the granularity of transmitted data.

**Simplicity**

Simplicity or development effort is defined as the effort that is needed to understand, design, implement, and re-engineer a web application. It is an important factor for the usage and acceptance of any new approach.

**Scalability**

In distributed environments scalability is defined by the degree of a systems ability to handle growing number of components. In Web engineering, a system's scalability is determined, for instance, by the degree in which a client can be served by different servers without affecting the results. A scalable Web architecture can be easily configured to serve a growing number of client requests.

**Portability**

Software that can be used in different environments is said to be portable. On the Web, being able to use the Web browser without the need for any extra actions required from the user, e.g., downloading plug-ins, induces the property of portability.

**Visibility**

Visibility [11] is determined by the degree in which an external mediator is able to understand the interactions between two components. The easier it is for the mediator to understand the interactions, the more visible is the interaction between those two components. Looking at the current implementations of AJAX frameworks, visibility in the client/server interactions is poor, as they are based on proprietary protocols.

## 4.2. Architectural Elements

Following [11, 23] the key architectural elements of SPIAR are divided into three categories, namely processing, data, and connecting elements. An overview of the elements is depicted in Figure 1, which will be explained in Section 4.3.

**Processing Elements**

The processing elements are defined as those components that supply the transformation on the data elements.

The *Client Browser* offers support for a set of standards such as HTTP, HTML, Cascading Style Sheets, JavaScript, and Document Object Model. It processes the representational model of a web page to produce the user interface. The user interaction can be based on a single page user interface model. All the visual transitions and effects are presented to the user through this interface. Just like a desktop client application, it consists of a single main page with a set of identifiable widgets. The properties of widgets can be manipulated individually while changes are made in-place without requiring a page refresh.

The AJAX *Engine* is a client engine that loads and runs in the client browser. There is no need for a plug-in for the web application to function. However, downloading the engine does introduce an initial latency for the user which can be compensated by the smaller data transfers once the engine is in place. The engine is responsible for the initialization and manipulation of the representational model. It handles the events initiated by the user, communicates with the server, and has the ability to perform client-side processing.

The *Server Application* resides on the server and operates by accepting HTTP-based requests from the network, and providing responses to the requester. All server-side functionality resides in the server application processing element.

The *Service Provider* represents the logic engine of the server and processes state changes and user requested actions. It is capable of accessing any resource (e.g., database, Web Services) needed to carry out its action. A Service Provider's functionality is invoked by event listeners, attached to components, initiated by incoming requests.

The *Delta Encoder/Decoder* processes outgoing/incoming delta messages. It is at this point that the communication protocol between the client and the server is defined and hidden behind an interface. This element supports delta communication between client and server which improves user-perceived latency and network performance.

*UI Components* consist of a set of server-side UI components. The component model on the server is capable of rendering the representational model on the client. Each server-side component contains the data and behavior of that part of the corresponding client-side widget which is relevant for state changes; There are different approaches as when and how to render the client-side UI code. GWT, for instance, renders the entire client-side UI code compile-time from the server-side Java components. Echo2, on the other hand, renders the components at run-time and keeps a tree of components on both client and server side. These UI components have event listeners that can be attached to client-side user initiated events such as clicking on a button. This element enhances simplicity by providing off-the-shelf components

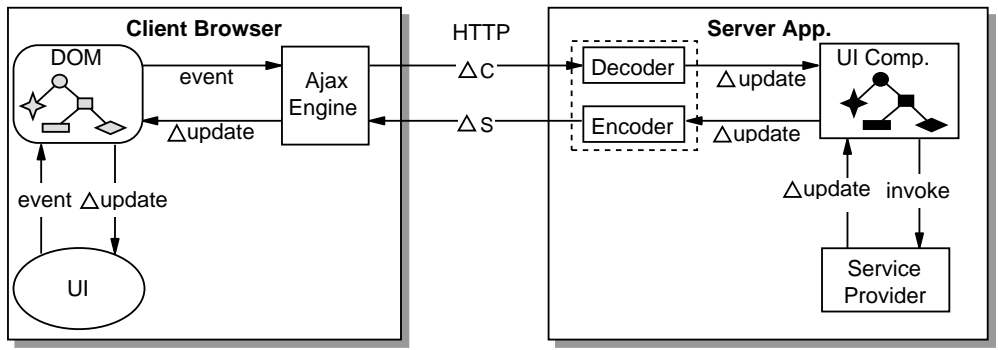

**Figure 1. Processing View of a** SPIAR**-based architecture.**

to build web applications.

**Data Elements**
The data elements contain the information that is used and transformed by the processing elements.

The *Representation* element consists of any media type just like in REST. HTML, CSS, and images are all members of this data element.

The *Representational Model* is a run-time abstraction of how a UI is represented on the client browser. The Document Object Model inside the browser has gained a very important role in AJAX applications. It is through dynamically manipulating this representational model that rich effects have been made possible. Some frameworks such as Backbase use a domain-specific language to declaratively define the structure and behaviour of the representational model. Others like GWT use a direct approach by utilizing JavaScript.

*Delta communicating messages* form the means of the delta communication protocol between client and server. SPIAR makes a distinction between the client delta data (DELTA-CLIENT) and the server delta data (DELTA-SERVER). The former is created by the client to represent the client-side state changes and the corresponding actions causing those changes, while the latter is the response of the server as a result of those actions on the server components. The delta communicating data are found in a variety of formats in the current frameworks, e.g., XML, BXML, JavaScript Object Notation (JSON), JavaScript. The client delta messages contain the needed information for the server to know for instance which action on which component has to be carried out. GWT uses an RPC style of calling services while in Backbase and Echo2 a component-based approach is implemented to invoke event listeners.

**Connecting Elements**
The connecting elements serve as the glue that holds the components together by enabling them to communicate.

*Events* form the basis of the interaction model in SPIAR. An event is initiated by each action of the user on the interface, which propagates to the engine. Depending on the type of the event, a request to the server, or a partial update of the interface might be needed. The event can be handled asynchronously, if desired, in which case the control is immediately returned to the user.

On the server, the request initiated by an event *invokes* a service. The service can be either invoked directly or through the corresponding UI component's event listeners.

*Delta connectors* are light-weight communication media connecting the engine and the server using a request/response mechanism over HTTP.

*Delta updates* are used to update the representational model on the client and the component model on the server to reflect the state changes. While a delta update of the representational model results in a direct apparent result on the user interface, an update of the component model invokes the appropriate listeners. These updates are usually through procedural invocations of methods.

### 4.3. Processing View

Given the processing, data, and connecting elements, we can use different architectural views to describe how the elements work together to form an architecture. Here we use a processing view, which concentrates on the data flow and some aspects of the connections among the processing elements with respect to the data [11]. This view is fits in the Components and Connectors viewtype as discussed by [8].

Figure 1 depicts the processing view of an SPIAR-based architecture based on run-time components rendering as in, e.g., Echo2. The view shows the interaction of the different components some time after the initial page request (the engine is running on the client). User activity on the user interface fires off an event to indicate some kind of component-defined action which is delegated to the AJAX engine. If a listener on a server-side component has registered itself with the event, the engine will make a DELTA-CLIENT message of the current state changes with the corresponding events and send it to the server. On the server, the decoder will convert the message, and identify and notify the relevant compo-nents in the component tree. The changed components will ultimately invoke the event listeners of the service provider. The service provider, after handling the actions, will update the corresponding components with the new state which will be rendered by the encoder. The rendered DELTA-SERVER message is then sent back to the engine which will be used to update the representational model and eventually the interface. The engine has also the ability to update the representational model directly after an event, if no round-trip to the server is required.

## 4.4. Architectural Constraints

Architectural constraints can be used as restrictions on the roles of the architectural elements to induce the architectural properties desired of a system. Table 1 presents an overview of the constraints and induced properties. A "+" marks a direct positive effect, whereas a "–" indicates a direct negative effect.

SPIAR rests upon the following constraints chosen to retain the properties identified previously in this section.

**Interaction and Synchronicity**
The client-server interaction can be realized in both a push- or pull-based style. In a push-based style [15], the server broadcasts the state changes to the clients asynchronously every time its state changes. Event-based Integration [2] and Asynchronous REST [16] are event-based styles allowing asynchronous notification of state changes by the server. This style of interaction is mainly supported in peer-to-peer architectural environments.

In a pull-based style, client components actively request state changes. Event-driven [22] architectures are found in distributed applications that require asynchronous communication, for instance, a desktop application, where user initiated UI inputs serve as the events that activate a process.

AJAX applications are designed to have a high user interactivity and a low user-perceived latency. *Asynchronous interaction* allows the user to, subsequently, initiate a request to the server at any time, and receive the control back from the client instantly. The requests are handled by the client at the background and the interface is updated according to server responses. This model of interaction is substantially different from the classic synchronous request, wait for response, and continue model.

**Delta-communication**
Redundant data transfer which is mainly attributed to retransmissions of unchanged pages is one of the limitations of classic web applications. Many techniques such as caching, proxy servers and fragment-based resource change estimation and reduction [5], have been adopted in order to reduce data redundancy. Delta-encoding [19] uses caching techniques to reduce network traffic, however, it does not reduce

**Table 1. Constraints and induced properties**

| | User Interactivity | User-perceived Latency | Network Performance | Simplicity | Scalability | Portability | Visibility |
|---|---|---|---|---|---|---|---|
| Asynchronous Interaction | + | + | | | | | |
| Delta Communication | + | + | + | | – | | – |
| Client-side processing | + | + | + | | | | |
| UI Component-based | + | | | + | | | |
| Web standards-based | | | | + | | + | |
| Stateful | + | + | + | | – | | – |

the computational load since the server generates the entire page for each request [21].

SPIAR goes one step further, and uses a *delta-communication* style of interaction. Here merely the state changes are interchanged between the client and the server as opposed to the full-page retrieval approach in classic web applications. Delta-communication is based on delta-encoding architectural principles but is different: delta-communication does not rely on caching and as a result, the client only needs to process the deltas. All AJAX frameworks hide the delta-communication details from the developers.

This constraint induces the properties of network performance directly and as a consequence user-perceived latency and user interactivity. Network performance is improved because there are less redundant data (merely the delta) being transported.

**User Interface Component-based**
SPIAR relies on a single page user interface with components similar to that of desktop applications, e.g., AWT's UI component model. This model defines the state and behavior of UI components and the way they can interact.

UI component programming improves simplicity because developers can use reusable components to assemble a Web page either declaratively or programmatically. User interactivity is improved because the user can interact with the application on a component level, similar to desktop applications.

**Web standards-based**
Constraining the Web elements to a set of standardized formats is one way of inducing portability on the Web. This constraint excludes approaches that need extra functionality (e.g., plug-ins, virtual machine) to run on the Web browser, such as Flash and Java applets, and makes the client cross-browser compatible. This constraint limits the nature of the data elements to those that are supported by web browsers.

**Client-side Processing**
Client-side processing improves user interactivity and user-perceived latency through round-trip reduction. For instance, client-side form validation reduces unnecessary server-side error reports and reentry messages. Additionally, some server-side processing (e.g., sorting items) can be off-loaded to clients using mobile code that will improve server performance and increase the availability to more simultaneous connections. As a tradeoff, client performance can become an issue if many widgets need processing resources on the client. GWT takes advantage of client-side processing to the fullest, by generating all the UI client-side code as JavaScript and run it on the client.

**Stateful**
A stateless server is one which treats each request as an independent transaction, unrelated to any previous request, i.e., each request must contain all of the information necessary to understand it, and cannot take advantage of any stored context on the server [12]. Even though the Web architecture and HTTP are designed to be stateless, it is difficult to think of stateless Web applications. Within a Web application, the order of interactions is relevant, making interactions depend on each other, which requires an awareness of the overall component topology. The statefulness is imitated by a combination of HTTP, client-side cookies, and server-side session management.

Unlike REST, SPIAR does not constrain the nature of the state explicitly. Nevertheless, since a stateless approach may decrease network performance (by increasing the repetitive data), and because of the component-based nature of the user interactions, a stateful solution might become favorable at the cost of scalability and visibility.

## 5. Discussion

In this section we use SPIAR to discuss various decisions and tradeoffs to be made when developing AJAX frameworks and applications.

**Resource-based versus Component-based**
The architecture of the World Wide Web [31] is based on resources identified by Uniform Resource Identifiers (URI), and on the protocols that support the interaction between agents and resources. Using a generic interface and providing identification that is common across the Web for resources has been one of the key success factors of the Web.

The nature of Web architecture which deals with Web pages as resources causes redundant data transfers [5]. The delta-communication way of interaction in SPIAR is based on the component level and does not comply with the Resource/URI constraint of the Web architecture. The question is whether this choice is justifiable. To be able to answer this question we need to take a look at the nature of interactions within single page applications: safe versus unsafe interactions.

**Safe versus Unsafe Interactions**
Generally, client/server interactions in a Web application can be divided into two categories of *Safe* and *Unsafe* interactions [30]. A safe interaction is one where the user is not to be held accountable for the result of the interaction, e.g., simple queries. An unsafe interaction is one where a user request has the potential to change the state of the resource.

In single page Internet applications, where interaction becomes more and more desktop-like, where eventually *Undo/Redo* replaces *Back/Forward*, the safe interactions remain using URIs while the unsafe ones can be *safely* carried out at the background using delta-communication in which neither the data transmitted nor the data received in the response necessarily correspond to any resource identified by a URI. This implies the engine should also provide the means of linking to safe operations as well as hyper-linked documents. The URI's *fragment identifier* can be used for this purpose. Interpretation of the fragment identifier is then performed by the engine that dereferences a URI to identify and represent a state of the application.

**Client- or server-side processing**
Within the current frameworks it is not possible for developers to choose whether some certain functionality should be processed on the client or on the server. How the computation is distributed can be an important factor in tunning a web application. AJAX frameworks architectures should provide the means for the developer to decide if and to what extend computation should be done on the client.

**Asynchronous Synchronization**
The asynchronous interaction in AJAX applications may cause race conditions if not implemented with care. The user can send a request to the server before a previous one has been responded. In a server processor that handles the requests in parallel, the second request can potentially be processed before the first one. This behavior could have drastic effects on the synchronization and state of the entire application. A possible solution would be handling the event-triggered requests for each client sequentially at the cost of server performance.

**Communication Protocol**
As we have seen, currently each AJAX framework has implemented its own specific communication protocol. This makes the visibility of client/server interactions poor as one must know the exact protocol to be able to make sense of the delta messages. It also results in a low level of scalability for these applications. For a client to be able to communicate with an AJAX server, again it needs to know the protocol of that server application. These two properties can be improved by defining a standard protocol specification for the communication by and for the AJAX community.

**Design Models**
Figure 2 shows a meta-model of an AJAX web application. The UI is composed of widgets of UI components. The client single page is built by the server-side widgets. Delta changes as well as view changes occur on the widget level. A view change, can be seen as navigating through the available widgets. AJAX frameworks should provide clear navigational models for developers. Research is needed to propose design models for AJAX developers by for instance extending the UML language to model user interaction, navigation through components, asynchronous/synchronous actions and client versus server side processing.

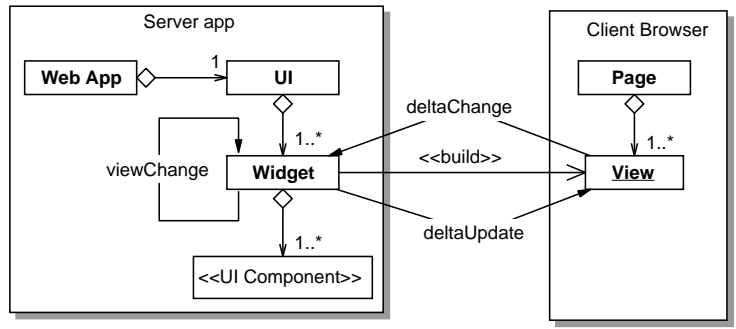

**Figure 2. A single page web application composed of UI components.**

**Fragment-based Approach**
The page-sequence model of the Web makes it difficult to treat portions of Web pages (fragments), independently. Fragment-based research [5, 6, 7] aims at providing mechanisms to efficiently assemble a Web page from different parts to be able to cache the fragments. Recently proposed approaches include several server-side and cache-side mechanisms. Server-side techniques aim at reducing the load on the server by allowing reuse of previously generated content to serve user requests. Cache-side techniques attempt to reduce the latency by moving some functionality to the edge of the network. These fragment-based techniques can improve network and server performance, and user-perceived latency by allowing only the modified or new fragments to be retrieved.

Although the fragments can be retrieved independently, these techniques lack the user interface component interactivity required in interactive applications. The UI component-based model of the SPIAR style in conjunction with its delta-communication provides a means for a client-/server interaction based on state changes that does not rely on caching.

## 6. Related Work

While the attention for rich Internet applications in general and AJAX in particular in professional magazines and Internet technology related web sites has been overwhelming, few research papers have been published on the topic so far.

Recently a number of technical books have appeared on the subject of developing AJAX applications. Asleson and Schutta [1] focus primarily on the client side aspects of the technology and remain 'pretty agnostic' to the server side. Crane et al. [9] provide an in-depth presentation of AJAX web programming techniques and prescriptions for best practices with detailed discussions of relevant design patterns. They also mention improved user experience and reduced network latency by introducing asynchronous interactions as the main features of such applications. While these books focus mainly on the implementation issues, our work examines the architectural design decisions and properties from an abstraction level by focusing on the interactions between the different client/server components.

Pace [26] is an event-based architectural style for trust management in decentralized applications. TIGRA [10] is a distributed system style for integrating front-office systems with middle- and back-office applications. Aura [25], an architectural framework for user mobility in ubiquitous environments, uses models of user tasks as first class entities to set up, monitor and adapt computing environments.

Khare and Taylor [16] evaluate and extend REST for decentralized settings and represent an event-based architectural style called ARRESTED. The asynchronous extension of REST, called A+REST, permits a server to broadcast notifications of its state changes to 'watchers'. This work is highly related to the concepts of AJAX applications. Applying a real push-based interaction style to AJAX, however, will probably take some time as the standard browsers and servers do not support this form of communication yet.

The SPIAR style itself draws from many existing styles [16, 22, 24, 27] and software fields [11, 19, 23], discussed and referenced in the paper. Our work relates closely to the software engineering principles of the REST style [12]. While REST deals with the architecture of the Web [31] as a whole, SPIAR focuses on the specific architectural decisions of AJAX frameworks.

## 7. Concluding Remarks

In this paper we have discussed SPIAR, an architectural style for AJAX. The contributions of this paper are in two research fields: web application development and software architecture.

From a software architecture perspective, our contribution consists of the use of concepts and methodologies obtained from software architecture research in the setting of AJAX Internet applications. Our paper further illustrates how the architectural concepts such as properties, constraints, and different types of architectural elements can help to organize and understand a complex and dynamic field such as single page Internet development. In order to do this, our paper builds upon the foundations offered by the REST style, and offers a further analysis of this style for the purpose of building web applications with rich user interactivity.

From a web engineering perspective, our contribution consists of the SPIAR style itself, which captures the guiding software engineering principles that practitioners can use when constructing and analyzing AJAX frameworks as well as applications. The style is based on an analysis of various of such frameworks, and we have used it to address various design tradeoffs and open issues in AJAX applications.

Future work encompasses the use of SPIAR to analyze and influence AJAX developments. One route we foresee is the extension of SPIAR to incorporate additional models for representing, e.g., navigation or UI components, thus making it possible to adopt a model-driven approach to AJAX development. At the time of writing, we are using SPIAR in the context of enriching existing web applications with AJAX capabilities.

**Acknowledgments** Partial support was received from SenterNovem, project Single Page Computer Interaction (SPCI). We thank Bas Graaf (TU Delft), Tijs van der Storm (CWI), and Mark Schieffelbein (Backbase) for their feedback on our paper. We particularly would like to thank Kees Broenink (Backbase) for our earlier collaboration on SPIAR [18].

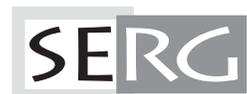